# Is the Lorentz Transformation Distance-Dependent ?

Ernst Karl Kunst

**An analysis of the Lorentz transformation shows that the unchangeability of the space-time coordinates of the inertial systems under consideration and the possibility of a direct projection of those coordinates onto another are the underlying basic assumptions as to its unlimited validity. It is demonstrated that from an empiric-physical point of view these assumptions are not given in the case of inertial systems separated by very large distances. Analogous to the impossibility to measure motion relative to absolute space, it turns out to be physically non feasible to extend the coordinate system of any reference frame considered at rest relative to a distantly moving system for a direct comparison of the coordinates, and vice versa.**
**The extended Lorentz transformation strictly based on first physical principles predicts the possibility of superluminal propagation of very distantly moving material bodies and, in this case, the generation of Cerenkov radiation out of the quantum vacuum. For many astrophysical phenomena and their experimentally verified properties this yields a novel view.**

## Introduction

In the last decades superluminal motions in extragalactic radio and optical sources on the grounds of the validity of the Hubble-relation for cosmic distances, and recently also in the Milky Way, have been discovered. Furthermore, astronomers have been observing intra day and even shorter variability in luminosity especially in lacertae objects, nuclei of galaxies, quasars and jets of quasars as highly energetic phenomena. The most widely discussed explanation for those phenomena are astrophysical "beaming models" or "Doppler boosting", due to the orientation of a moving object toward the observer at a small angle to the line of sight, presuming the Lorentz transformation and, therewith, special relativistic aberration to be valid over huge cosmic distances relative to our vantage point on Earth [1].

In the following analysis of inertial motion is shown that the Lorentz transformation is distant-dependent with the most important result, superluminal motion of material bodies in principle to be possible. Starting point of this review is the validity of special relativity at close range or negligable separation of the inertially moving systems under consideration, though we refer to that theory in its symmetric form [2], which we recapitulate here in short. In the mentioned work on relativistic kinematics has been shown a preferred rest frame of nature ($\Sigma_0$) in any inertial motion to exist and any velocity ($v_0$) to be symmetrically composite or quantized. From this a symmetric modification of the Lorentz transformation follows between a frame of reference $S_1$ considered to be at rest according to the principle of relativity and a moving frame $S_2$:

$$x_2' = \gamma_0(x_1 - v_0 t_1), \qquad y_2' = y_1, \qquad z_2' = z_1, \qquad t_2' = \gamma_0(t_1 - \frac{v_0}{c^2} x_1),$$

$$x_1^\circ = \gamma_0(x_2' + v_0 t_2'), \qquad y_1^\circ = y_2', \qquad z_1^\circ = z_2', \qquad t_1^\circ = \gamma_0(t_2' + \frac{v_0}{c^2} x_2'),$$

(1)

where

$$\gamma_0 = \left(1 - \frac{v_0^2}{c^2}\right)^{-\frac{1}{2}}.$$

The dashed symbols in (1) designate the moving system $S_2$ and the open circles the system $S_1$ at rest, now considered moving relative to $\Sigma_0$ and $S_2'$. Likewise the observer resting in $S_2$ will deduce the respective transformation, which we do not cite here. Furthermore has been shown to be valid:

$$\begin{aligned} x_2 = x_1, \qquad & y_2 = y_1, \qquad z_2 = z_1, \qquad t_2 = t_1, \\ x_1' = x_2', \qquad & y_1' = y_2', \qquad z_1' = z_2', \qquad t_1' = t_2', \\ x_2^\circ = x_1^\circ, \qquad & y_2^\circ = y_1^\circ, \qquad z_2^\circ = z_1^\circ, \qquad t_2^\circ = t_1^\circ \end{aligned}$$

(1a)

and always $|v_0| = |-v_0|$. If into the second lines of (1) the upper lines are inserted, the identity results:

$$\begin{aligned} x_1^\circ \equiv x_1, \qquad & t_1^\circ \equiv t_1, \\ x_2^\circ \equiv x_2, \qquad & t_2^\circ \equiv t_2. \end{aligned}$$

(1b)

### 1. On very Distant Measures and Motions

As a physical basis in special relativity and also in its symmetric form the Lorentz transformation is derivated by taking the dimensional axes of the systems under consideration always to coincide and to be parallel, or with other words: It is taken for guaranted that the space-time coordinates of an event in the moving frame are directly projectable onto the respective ones of the reference frame considered to be at rest and vice versa [3]. The expression as laws in Minkowskian four-dimensional space-time [4] or its formulation as index-calculus (four-vector calculus) do not alter the underlying physical principle. Thus, any statement about relative motion in the framework of relativistic kinematics basically implies velocity of a moving system or body within or at very close range of the space-time coordinates of the frame of the observer, considered resting. This implies that, empirically speaking, it is unjustified to apply the transformation automatically on the kinematical relations between any two bodies of reference, which are separated by a considerable (e. g. interplanetary, interstellar or cosmic) spatial distance, because this proceeding obviously requires the space-time coordinates of the frame of reference based on the volume of the body "at rest" to be continued to the "moving" body or volume of reference and vice versa. If we try to extend the frame of reference far beyond the limits of the physical body resting in its coordinate source, we leave secure physic-empirical experience. In the contrary, we are forced to admit that a concrete elongation of the coordinate system "at rest" to the moving one (and vice versa) cannot be physically realized. We only could generate a new coordinate system near the "moving" one, very far though resting relative to the original system at rest. Just as it proves impossible to measure velocity relative to Newtonian "absolute space", a direct measurement of the dimensions and the velocity of a very distantly moving body resting in the source of its inertial system relative to an imaginary extension of the inertial frame at rest, turns out physically to be non feasible. Instead, the very physical basis of statements about the parallel translational motion of a distant object are solely measurements of light signals from that object in reference to an unit of measure of the own volume of reference, considered to be at rest in accord with the principle of relativity.

$S_1$ and $S_2'$ may be frames of reference with all coordinate axes parallel, which for the transformation equations (1) are valid. Especially may $S_2'$ be in relative motion at close range or within the rest frame of reference $S_1$ in the direction of the x-axis of the latter system. Very distantly from both systems we introduce a system $S_1^*$ in space, resting relative to $S_1$ so that $\overline{S_1^* S_2'} \gg \overline{S_1 S_2'}$, also all coordinate axes being parallel to the axes of $S_1$ and $S_2'$. The velocity of a light signal propagating through vacuum in the direction of the x-axis of $S_2'$ must be "c" for an observer resting in the latter system as well as in the nearby system $S_1$.

Now let us turn toward the very distant system $S_1^*$. For an observer, resting there, the velocity of light in $S_2'$ will appear apparently slowed in dependence on the distance $\overline{S_1^* S_2'}$ as well as the transversal motion of $S_2'$ relative to $S_1^*$.

Thus, the apparent or relative transversal velocity of $S_2'$ and the velocity of light within this system tend for an observer in $S_1^*$ toward null if he is only sufficiently apart from $S_2'$. Obviously the adjective "apparent" does not quite correspond to the facts because - as already stated - in physical reality there are no other light signals available for measurements of measures and velocities than those received from $S_2'$. Any statements as to the true velocity of $S_2'$ and the true velocity of light in that system prove meaningless for real measurements in the coordinate system $S_1^*$. Besides the relativity of inertial motion are dimensions, velocities and the velocity of light of distant bodies further relativated by the spatial distance between the inertial systems under consideration. These shrinked dimensions and retarded motions are the only empirically ascertainable ones and must therefore considered integral part of the principle of relativity. Thus, as the basic physical principles of relativistic kinematics of bodies, moving very far away from each other and, therewith, of the transformation equations, only four statements supported by physical experience exist:

1) Validity of the (symmetric) Lorentz transformation at negligable (nearly zero) distance between the inertial systems under consideration;

2) Physical impossibility to extend the coordinates of the reference frame "at rest" to a very distantly moving object and its coordinate system (and vice versa) and to measure dimensions and velocities by direct comparison of coordinates;

3) Apparent decrease of the velocity of distant objects, moving transverse to the line of sight and their dimensions, whereby the "apparent velocity" as well as the shrinkage of dimensions is a simple function of the respective value at the imaginary distance null and the distance in Minkowskian space-time between the bodies under consideration;

4) Independence of the dilation of time of moving objects of their distance and of the direction of the vector of velocity relative to the observer considered to be at rest.

The Lorentz transformation of a complete kinematic theory must be in accord with these first physical principles.

## 2. The Symmetric Lorentz Transformation between Systems Separated by Large Distances

It is obvious that the distance $R = \overline{S_1^* S_2^'}$, as observed from $S_1^*$, must directly join to the coordinate source (coincident with the center of gravity) of the latter system so that the velocity of $S_2^'$ relative to $S_1^*$ and the velocity of light at $S_2^'$ is slowed in accordance with proposition 3) and that at the distance $R \to 0$ proposition 1) and, therewith, (1) becomes fully valid. On the other hand the observer at $S_1^*$ according to proposition 4) continues to observe the lapse of time of $S_2^'$ being retarded. This also can be proven directly. Considering a Feynmanian "light signal watch", resting relative to $S_2^'$, we find that the distance factor R cancels out so that $t_2^' = \gamma_0 t_1^*$.

The γ-factor of the x-coordinate can be directly computed from (1), whereby we write $x_1 = x_1^*$, $x_1^° = x_1^{°*}$, $t_1 = t_1^*$ and $t_1^° = t_1^{°*}$:

$$\frac{x_2^'}{R} = \gamma \left( x_1^* - \frac{v_0 t_1^*}{R} \right), \quad \frac{x_1^{°*}}{R} = \gamma \left( x_2^' + \frac{v_0 t_2^'}{R} \right), \tag{2}$$

The only empirically known and certain physical experience is the shrinkage of the projection of the dimensions of a moving body plus the distance covered by it in a unit of time of the system at rest according to proposition 3), which facts are expressed by (2). By multiplying both equations (2), whereby according to (1a) and (1b) it is clear that $x_1^° = x_1$ and $t_1^° = t_1$ and, therewith, $x_1^{°*} = x_1^*$ and $t_1^{°*} = t_1^*$, we receive

$$Y = \frac{1}{R\sqrt{1 - \dfrac{v_0^2}{c^2 R^2}}}$$

if always $x = ct$. Inserting this expression into (2) delivers the far range transformation of relativistic kinematics and its inverse

$$x_2' = Y_0^* \left( x_1^* - \frac{v_0 t_1^*}{R} \right), \quad y_2' = y_1^*, \quad z_2' = z_1^*, \quad t_2' = Y_0^* \left( t_1^* - \frac{v_0 x_1^*}{c^2 R} \right),$$

$$x_1^{\circ*} = Y_0^* \left( x_2' + \frac{v_0 t_2'}{R} \right), \quad y_1^{\circ*} = y_2', \quad z_1^{\circ*} = z_2', \quad t_1^{\circ*} = Y_0^* \left( t_2' + \frac{v_0 x_2'}{c^2 R} \right),$$

(3)

where

$$Y_0^* = \frac{1}{\sqrt{1 - \dfrac{v_0^2}{c^2 R^2}}}, \quad Y_0 = \frac{1}{\sqrt{1 - \dfrac{v_0^2}{c^2}}}.$$

Equations (3) are in full accord with the four basic physical propositions for inertially moving systems well separated in space by a considerable distance and, therefore, govern the transformation of their event coordinates. At interplanetary, interstellar and intergalactic distances this is the normal case. Thus, the original Lorentz transformation - though in its symmetric form (1) - proves to be a border line case at close range with mere "local" validity and it is obvious that (3) is the far-range form of (1), passing into the latter if $R \to 0$ and $Y_0^* = Y_0$, $x_1^* = x_1$, $x_1^{\circ*} = x^\circ$.

### 3. Modification of the Principle of Relativity in the Far-Range Case

From (3) is evident that

$$x_2' = x_1^*, \quad t_2' = t_1^* Y_0$$

if $|R| \gg |c|$ and $Y_0^* = 1$ and hence the velocity of an object resting in $S_2'$, moving very far from the frame of reference $S_1^*$, must be

$$\frac{\Delta x_1^*}{\Delta t_1^*} = \frac{\Delta x_2'}{\Delta t_2'} \gamma_0$$

which becomes to

$$u_{x_1^*} = u_{x_2'} \gamma_0 = v_0 \gamma_0. \tag{4}$$

This result is also supported by the following consideration:
For the relative movement of the systems $S_2'$ and $S_1^*$, separated by a considerable distance, must according to (3) and if $|R| \gg |c|$ be valid:

$$\int_{t_1=0}^{1} v_{0_1} dt_1 = \int_{t_2=0}^{1} v_{0_2} dt_2 \tag{5}$$

- the dashes and asterisks of (3) are abandoned in favour of a simplified notation - and because of the absolute symmetry relative to $\Sigma_0$ also the proper time integral (eigenzeit) must have the same numerical value in both frames of reference. But because the elapsed time differs, especially as observed from $S_1$, the time particle $dt_2$ seems expanded by the value

$$dt_2 = dt_1 \gamma_{0_1},$$

from (5) follows:

$$v_{0_1} = v_{0_2} \gamma_{0_1},$$

where $v_{01}$ means velocity of $S_2$ relative to $S_1$. Because (of its composite nature) $v_{02} = 2v_1/(1 + v_1^2/c^2)$, in any case must be $v_{02} > v_1$, implying $v_{02} = v_0$. Introducing the symbol "$V_0$" for the velocity $v_{01}$ this results in connection with (4) in

$$V_0 = v_{0_1} = u_{x_1^*} = v_0 \gamma_0, \tag{6}$$

Ultimately we have

$$V_0 \int_{t_1=0}^{1} dt_1 = v_0 \int_{t_2=0}^{1} dt_2. \tag{7}$$

Let an inertial system $S_0$ be at rest relative to the cosmic microwave background so that it presumably also rests relative to space-time. Now, suppose two systems $S_1$ and $S_2$ to move relative to $S_0$ at equal but oppositely directed velocity $v_0$ so that in the case where (3) is valid we have $x_2 = x_1 = x_0$ and $t_2 = t_1 = t_0 \gamma_0$, wherefrom follows

$$\frac{\Delta x_1}{\Delta t_1} = \frac{\Delta x_2}{\Delta t_2} < \frac{\Delta x_0}{\Delta t_0} \to v_1 = v_2 < V_0 = \text{Max.} \qquad (8)$$

in accordance with (7). On the other hand, for an obserer based at $S_1$ according to special relativity should also be valid

$$\frac{\Delta x_2}{\Delta t_2} < \frac{\Delta x_0}{\Delta t_0} < \frac{\Delta x_1}{\Delta t_1} \to v_2 < v_0 < V_1 = \text{Max.} \qquad (8a)$$

Evidently (8a) contradicts (8). This contradiction can obviously only be resolved if all motions are related to a system at rest relative to space-time. Otherwise we would, as (8a) shows, arrive at contradictory results. Hence (8) must in any case be true. This implies that also in the far-range case the scalar

$$x_1^2 + y_1^2 + z_1^2 - c_1^2 t_1^2 = x_0^2 + y_0^2 + z_0^2 - c_0^2 t_0^2$$

remains valid, though $c_1 \neq c_0 = c_1 \gamma_0$ so that $c_1 t_1 = c_0 t_0$. Thus, the state of motion of any inertial system relative to space-time can be expressed best and shortest by

$$|n| = \gamma_0,$$

where $n(\beta_0', 1)$ is a complex number in the complex ct, x-plane of symmetrically modified Minkowskian space-time [2]. The twin-paradox of special relativity is resolved to the result that time dilation depends on motion relative to space time. Hence clocks on the Earth (in the whole system of the sun), which moves at velocity $v_0 \approx 600$ km/s relative to space-time, should run slow $\approx 2 \times 10^{-6}$ s, as compared with a clock at rest relative to the latter. This effect implies that spacecraft on a direct track to the outer planets, as e. g. Jupiter, should arrive there minutes earlier in dependence on the duration of the voyage.

### 4. Further Kinematic Consequences

We turn to the physical implications of (3) and its derivations (4) and (7). Evidently the dilation of time in $S_2'$ is compensated for by the symmetric inertial velocity

$$V_0 = u_{x_1^*} = \gamma_0 v_0$$

of the latter system relative to $S_1^*$ if both systems are far away from each other so that observers, resting in the coordinate sources of either system, will meet after the same amount of time has elapsed, as measured in their systems.

Possible effects of acceleration are neglected and it is understood that asymmetric ageing should occur if $S_2'$ is accelerated, as proven by general relativity. Nevertheless, in the far-range case any velocity even exeeding that of light in any amount is possible, allowing in principle the superluminal propagation of solid bodies and thereby transfer of information.

Consider ultra relativistic particles or photons to move relative to the very distant system $S_2'$ so that (3) is valid, according to the equations

$$x_2' = u_{x_2'} t_2', \quad y_2' = u_{y_2'} t_2', \quad z_2' = 0.$$

Transformation into the coordinates and the time of $S_1^*$ yields

$$u_{x_1^*} = u_{x_2'} \gamma_0, \quad u_{y_1^*} = u_{y_2'} \gamma_0, \quad u_{z_1^*} = 0,$$

wherefrom follows

$$\tan \Theta_1^* = \tan \Theta_2'.$$

Thus, no relativistic aberration or Doppler boosting is to observe by an observer at $S_1^*$, which implies that this special relativistic effect at close range is ruled out in the far-range case (3) as an explanation of superluminal phenomana on and in cosmic objects, especially jets.

It clearly follows that the principle of cause and effect is not impaired if $V_0 > c$.

### 5. Physical Effects to Expect from the Superluminal Propagation of very Distant Material Bodies

Analogous to the Cerenkov relation

$$\cos \alpha = \frac{ct}{n} \times \frac{1}{\beta ct} = \frac{1}{\beta n} \tag{9}$$

for electrical non neutral particles moving through a material medium at superluminal velocity $v = \beta c > c/n$, where "n" means the refractive index of the medium and "α" the half angle of the cone of radiation [5], we have to expect an electromagnetic shock-wave phenomenon, when a very distant material body, e. g. a particle, traverses vacuo at the velocity $V_0 = \gamma_0 v_0 \geq c$. This results from the fact that the probability "p" to encounter virtual photons (or elementary dipoles) for a particle traversing the fluctuating quantum vacuum at subluminal symmetric velocity $V_0 < c$ should according to Heisenberg's uncertainty relation $\Delta E \Delta t \geq h$ rise with growing velocity. To express the uncertainty relation as a function of velocity we write

$$\frac{\Delta E \Delta t c^2}{V_0^2} \geq \frac{h V_0^2}{c^2} \qquad (10)$$

so that the left-hand side attains the highest and the right-hand side the lowest possible value, where h is Planck's constant. This follows, because $\sqrt{h} = 2\lambda_1 = 2\tau_1 c = mc$, where $\lambda_1$ means fundamental length and $\tau_1$ quantum of time [6]. From (10) is derived

$$p = \frac{\Delta E \Delta t}{h} = \frac{\Delta E}{h \Delta \nu} \geq \frac{V_0^4}{c^4} \leq 1, \qquad (11)$$

where $V_0 \leq c$. Evidently (11) results in p = 1 if $V_0/c = 1$. This implies the energy E of virtual photons to become real and stable for a moving particle if its far-range velocity $V_0 = c$. In this case (11) delivers $\Delta E = h \Delta \nu$ and $\Delta \nu = \nu = 1$, because $\nu$ must attain the value of the lowest possible frequency. Hence if $V_0$ exceeds the velocity of light, from (11) follows

$$p = \frac{h V_0^4}{E c^4} = 1 = \text{const}, \qquad (12)$$

where always $V_0/c \geq 1$. Thus, real radiation arises off the vacuum and analogous to (9) we have

$$\cos \alpha = \frac{c}{V_0} = \frac{1}{\gamma_0 \beta_0}, \qquad (13)$$

where $V_0 \geq c$ and $\beta_0 = v_0/c$.
The Cerenkov half angle $\alpha$ of vacuo tends to a maximum value if $\gamma_0 \to 0$ and to its minimum value if $\gamma_0 \to 1$. According to (12) between superluminal velocity $V_0 \geq c$ and highest frequency of the radiated photons the relation

$$\frac{\vec{V_0}}{c^4} = \frac{E}{h} = \nu, \qquad (14)$$

is valid, wherefrom in connection with (13) follows

$$\cos \alpha = \left(\frac{h}{E}\right)^{-\frac{1}{4}} = \nu^{-\frac{1}{4}}. \qquad (15)$$

"E" in the sense of symmetrically modified special relativity [2] means center-of-mass energy $E = (2E_{phot})^{1/2}$, where $E_{phot}$ means conventional photon energy.

## 6. Evidence of Vacuum Cerenkov Radiation in High Energetic Astrophysical Phenomena

From the foregoing is clear that in the far-range case relativistic beaming is excluded as an explanation of the characteristics of jets and especially of the often observed superluminal motions in jets. According to this distance-dependent extention of special relativistic kinematics is superluminal motion as natural as any sub-light velocity.

It is predicted that most if not all non-thermal emission of cosmic objects is due to vacuum Cerenkov radiation.
With the polarization of space (vacuum) by a superluminal particle and the radiation of photons must be connected a continous energy (velocity) loss of the moving particle, analogous to the stopping power owing to the density effect of the theory of Cerencov radiation [5]. This effect naturally explains the variation in the continuum emission of Active Galactic Nucleii (AGN) and jets, where years after fast outbursts in the optical region corresponding events at radio frequencies are found [7]. High energetic particles generating optical or vacuum Cerenkov radiation of still higher frequencies, e. g. in a jet, would according to (14) travel at a velocity of $\approx 10^4$ c and, therewith, cover the whole length of the jet of some kpc in a couple of years, gradually loosing energy and slowing down. Thus an intensity variation, e. g. due to a sudden rise of the particle number in the jet stream, would first occur in the emission of highest frequency and then in the mentioned time wander through the whole frequency band down to the radio hot spot. The experimentally found strong variability of the gamma-ray luminosity of blazars and its correlation with the far-infrared luminosity of the latter [8] (also in other AGN) can also be explained by sudden variations of a particle flux streaming randomly off the blazar core at diferent highly relativistic velocities. This picture is strongly supported by the HST view of a certain class of radio galaxies, which revealed non-thermal nuclear sources with a linear correlation between the radio and optical luminosities [9].

The relativistic vacuum Cerenkov effect also explains in a fully way other observed phenomena, optical and otherwise, on and in jets of high energetic extragalactic systems, especially the marked colour variation along jets from blue over red, near infrared to radio wavelengths, as for instance in the elliptical galaxy M87 and even the occurrence of X-ray emission, as in the jets of the quasar 3C 273 [10] and again M87 [11]. Furthermore, the frequently observed onesidedness of the jet phenomenon in the frame of our theory can be easily explained by two counter directed ultra relativistic jets with radiation cones according to (13) and (15), and the jet axes mildly inclined against the line of sight. In this case the Cerenkov radiation from the farer jet is always directed away from the observer to remain undetectable, except from stray particles. This effect also explains the relative faintness of the counter jet of HH 30 in the Milky Way [12].
Finally most recent HST observations of optical jets in radio galaxies have shown definite examples of two-sided optical hotspots and jets [13], which clearly rule out Doppler boosting, but are explained easily by vacuum Cerenkov radiation.